\documentstyle[preprint,aps]{revtex}
\begin{document}
\title{ Metastability of NbN in the ordered vacancy NbO phase}
\author{E.C. Ethridge, S.C. Erwin, W.E. Pickett}
\address{Complex Systems Theory Branch, Naval Research Laboratory,
Washington DC 20375}
\date{May 10, 1995}
\maketitle
\begin{abstract}
A metastable phase of NbN with superconducting T$_c$=16.4~K was reported
recently by Treece and collaborators.  The reported structure of the thin film
sample deviates from the rocksalt (B1) NbN structure with 25\% ordered
vacancies on each sublattice (space group Pm3m) and a lattice constant of
 4.442~\AA.  Using full potential electronic structure methods, we contrast
the electronic structure with that of B1 NbN.  The calculated energy,
1.00~eV/molecule higher than B1 NbN, and calculated lattice constant of
4.214~\AA~indicate that the new phase must be something other than the ordered
 stoichiometric Pm3m phase.
\end{abstract}

Transition metal nitrides and carbides possess a number of useful
electronic and
physicochemical properties such as metallicity and superconductivity, extreme
hardness and brittleness, as well as high melting points \cite{toth,gubanov}.
These attributes make them
attractive candidates for applications ranging from composite materials to low
temperature electronic devices.  Metastable phases of these compounds have
recently been implicated as an improvement upon existing materials,
showing  enhancements in their electronic properties \cite{gubanov}.
In the simplest of the metastable phases, the structures
deviate from their ideal rocksalt phase, with ordered vacancies on either the
metal or nonmetal sublattice or both.

Recently a new phase of NbN was reported in thin films fabricated on MgO
using pulsed laser deposition (PLD) \cite{treece}.
 The sample was shown to
demonstrate a superconducting critical temperature of
16.4~ K and a lattice parameter of 4.442~\AA, both somewhat larger than
accepted values for rocksalt structure NbN (16K; 4.378~\AA).
This new phase of NbN, shown in Fig.~1, deviates from its common B1
(rocksalt) structure with 25\% ordered vacancies on both sublattices,
which also
occurs for NbO.  The new phase has Pm3m symmetry, and consists of
nonpenetrating Nb octahedra and nonpenetrating Nb octahedra centered at
vacancy sites.  Both octahedra have four-fold planar coordination at
each  atomic site and six-fold ``coordination'' at the vacancy site.
While it is intuitive that this system should reflect behavior similar
to that found in electronic and structural analyses of metastable binary
carbides, nitrides, and oxides,  a detailed
theoretical analysis of this new phase of NbN has not yet been carried out.
This phase is almost 30\% less
dense than the rocksalt phase, which already suggests strong alterations
in the electronic structure.  It is the purpose of this paper to
characterize the electronic states of
this new phase of NbN and, in particular, to understand the effect of
 the ordered vacancies on the electronic states.  In the discussion
to follow, detailed comparisons will be made between the vacancy
and rocksalt structure, which will be referred to as
Nb$_3$N$_3$ and Nb$_4$N$_4$, respectively.  We also report
calculations of the equation of state of both phases that
appear to be incompatible with the new phase as identified.

Density functional electronic band calculations were carried out
for Nb$_3$N$_3$ and Nb$_4$N$_4$ using a basis set of linear combination
 of atomic
orbitals (LCAO).  The exchange-correlation potential was treated within a
 local-density approximation (LDA) as parametrized by Perdew and
 Zunger \cite{perdew}.  Details of
this method have been described elsewhere\cite{erwin}.  In each
calculation self-consistency was achieved using 20 special k points
in the irreducible Brillouin zone.  The density of states (DOS) was
computed by diagonalizing the Hamiltonian at 104 k points in the
irreducible zone, followed by Fourier interpolation to a denser
grid and integration using the linear tetrahedron method.
The basis functions were expanded in a set of
17 Gaussian exponents contracted into seven s-type,
five p-type, and four d-type functions for Nb, and 12 exponents
contracted into three s-type and two p-type functions for N, for a total
of 55 LCAOs per NbN unit.

The Nb$_4$N$_4$ and Nb$_3$N$_3$ bandstructures are shown in Fig.~2.
For purposes of comparison, both calculations were performed in a
simple cubic cell using the reported lattice constant of
the Pm3m phase.  Mulliken population analysis was used to
characterize the orbital percentage of each band in both structures
\cite{mulliken}.
Figure 2 also indicates the orbital character of each of the bands,
according to the criteria that the Mulliken orbital population of a state
exceeds 50\%.

In the Nb$_4$N$_4$ structure, we observe the type of behavior
one has come to expect from IVa and Va metal nitrides in the rocksalt
structure\cite{papa}. Since each atom is octahedrally coordinated
with six atoms of the opposite type, $\sigma$ bonding may occur
between the Nb and N via the $a_{1g}$, $t_{1u}$, and
$e_g$ orbitals while $t_{2g}$ orbitals comprise the nonbonding orbitals
with respect to the Nb-N bonding.  The bands near the Fermi level
E$_F$  are made up predominantly of Nb states.  The Nb bands have
mostly $t_{2g}$ symmetry below E$_F$ and are a mixture of
$e_g$ and  $t_{2g}$ symmetry above E$_F$. The N p manifold
lies below the Nb d bands in the energy range of -8.6 to -3.5~eV.  The
onset of the Nb p manifold lies above the Nb d near 6~eV.

In the Nb$_3$N$_3$ structure, the loss of six-fold coordination gives
rise to anisotropic bonding near the vacancy sites that is reflected
in a markedly different bandstructure.
Notable features are (1) reduction in the bandwidth by 40\% and
(2) the presence of a highly dispersive
band which crosses the Fermi level and peaks at the R point. This band,
referred to as a ``vacancy band'',  is similar to that observed in
the isostructural NbO vacancy phase \cite{wimmer,schulz}.
It has Nb d character below E$_F$ and changes to primarily
Nb p character as it crosses the Fermi energy.
Mulliken population analysis shows that this is the only region for
which Nb p states are present within the Nb d complex. The remaining
Nb p bands are energetically higher than the Nb d bands.  The Fermi level
 falls in an energy region where both N p and Nb d bands are occupied.
The mean energy of the N p band has increased markedly, relative
to Nb$_4$N$_4$, as is apparent in Figure 2.
These findings are consistent with trends in the vacancy
stabilization mechanisms of TiO,\cite{huisman} as well as NbO
\cite{wimmer,schulz}.

The total and partial DOS for Nb$_4$N$_4$ is shown in Figure 3.  The peak
5.4 eV below E$_F$ is dominated by N p states with some
admixture of Nb d.  At higher energies the N p DOS decreases
 rapidly and contributes very little to the DOS at the
Fermi level.  Above the Fermi level, a
secondary peak is present near 2 eV, which originates from the Nb d states.
The onset of this peak occurs below the Fermi level and gives rise to the
observed N(E$_F$) of 3.9 states/eV-sc cell with primarily Nb d
$t_{2g}$ character.

For the  Nb$_3$N$_3$ structure, also shown in Fig. 3, N(E$_F$) is
6.9~states/eV-sc cell, an enhancement relative to its rocksalt counterpart
  of 75\% per unit volume, or 135\% per NbN unit. This arises from a
dramatic redistribution of the spectral weight of the Nb d
bands, which results in a sharply peaked structure at the Fermi level.
Although the spectral weight
of the the N p bands is shifted to a higher energy these states
are less significant to the enhancement of the DOS at E$_F$.

It is instructive to contrast the proposed  Nb$_3$N$_3$ phase with the
isostructural Nb$_3$O$_3$.   In Nb$_3$O$_3$, the Fermi level falls
slightly above a pseudogap in the DOS and results in a low N(E$_F$).  The
presence of this minimum is considered to be an indication of the
stabilization brought about by the ordered vacancies.  Andersen and
Satpathy~\cite{andersen}
have shown that the directed d$^4$ orbitals introduced by Cotton and Haas~
\cite{cotton} constitute a basis for understanding the stability of
 Nb$_3$O$_3$: all bonding orbitals are filled and a single antibonding
orbital is occupied.  In  Nb$_3$N$_3$, two of the bonding orbitals
are unoccupied.  In addition, the Fermi level falls at a position of
large density of states that leaves the uppermost occupied states with
high band energy.  Both of these effects disfavor the stability of the
Pm3m phase for NbN.

To gain insight into the role of the vacancies, effective electron
configurations were computed for the constituents of both structures
and summarized in Table I.  The total (Mulliken) charge on each atom
is decomposed according to orbital and symmetry type.  The symmetry
decomposed charge distribution  of  Nb$_4$N$_4$ reflects
the full cubic symmetry of the structure.   For Nb$_4$N$_4$,
the total charge transferred from Nb to N is 0.83.  The Nb$_3$N$_3$
structure is slightly less ionic with a charge transfer of 0.71.
A comparison of the two phases shows that the difference in the ionic
charges arises from a reduction in charge transfer from the  Nb~p
to the N~p states.

It is useful to define a  local coordinate system
at each atom with the z-axis directed toward a vacancy site (i.e.
along the axis of the D$_{4h}$ point group).
There are two types of (100) planes. The Nb-rich plane, defined at
y=0.5 (or equivalently at x=0.5) in Fig.1, is made up of Nb atoms
next nearest neighbor (nnn) coordinated with Nb atoms.
In this plane, the symmetry equivalent Nb xz, yz orbitals are directed
at nnn Nb atoms.  In the N-rich plane, defined at
z=0 in Fig.1, Nb atoms are nearest neighbor (nn) coordinated to N atoms.
Niobium xy, x$^2$-y$^2$ orbitals are directed at nn N's in this plane and
Nb-N~ dp$\sigma$ bonding is maintained.
As shown in Table I, the charge on these Nb orbitals is
smaller than the out-of-plane Nb orbitals.  This  indicates that while
charge transfer occurs in this plane due to Nb-N~$\sigma$ bonding, the
presence of vacancy sites in adjacent planes reduces the charge
transfer in the perpendicular direction.

To ascertain the relative stability of the two phases of NbN we have
calculated the equation of state.  We have used the full potential
linearized augmented plane wave method (LAPW) \cite{singh,wei} with the
Vosko-Wilk-Nusair \cite{vosko}
parametrization of the exchange-correlation energy and potential in
the local density approximation.  For all calculations sphere radii
were taken to be R$_{Nb}$=2.30 a.u. for Nb and R$_N$=1.55 a.u. for N.  For
Nb$_3$N$_3$ empty spheres of the same sizes were placed at the Nb and
N vacancies respectively.  The basis set cutoff was defined by
R$_N$K$_{max}$=7.0, which translates to R$_{Nb}$K$_{max}$=10.4.  The basis
set size ranged from 760 LAPWs for the smallest volume to 1080 LAPWs for the
largest volumes, for the simple cubic cell that was used.  The Nb 3p
semicore states were treated in the same window for additional accuracy
that might be necessary at the smaller volumes considered, and additional
local functions were used for the Nb 3p and 4d states and for the N 2s
states.  Twenty special k points were used in all calculations.

The calculated energy, presented versus simple cubic lattice constant,
are shown for both phases in Figure~4.  There are two unexpected results:
(1) the predicted lattice constant for the Nb$_3$N$_3$ phase, 4.214~\AA,
is much smaller than the reported value [2] of 4.442~\AA, and (2)
the energy of the Nb$_3$N$_3$ phase is {\it 1.00 eV higher} per NbN
unit than in the rocksalt phase.  The calculated lattice constant
of the rocksalt phase is 4.365~\AA, within 0.3\% of the experimental
value, lending confidence to our predictions.

Other results from our equation of state include the bulk modulus B and
its pressure derivative B': for the Nb$_4$N$_4$ phase B=3.57~Mbar,
B'=4.54; for the  Nb$_3$N$_3$ phase B=3.84~Mbar, B'=4.38.
These equation of state results are not compatible with the identification
of the structure of the PLD NbN film.  First, the equilibrium lattice constant
is 5\% less than the reported one.  This could conceivably be rationalized
as a sufficiently thin film stabilized in a highly strained film due
to epitaxial registry, except that the MgO substrate has a lattice constant
(4.21-4.24~\AA~depending on temperature) that is 5\% smaller than the
reported value of 4.442~\AA.  Second, the energy
difference of 1.00 eV per NbN unit
is exceedingly large, and a 5\% strain would increase this difference
noticeably.  The stored energy in even a few monolayers of the Nb$_3$N$_3$
phase could never be sustained by bond formation at the interface.

In summary, we have found that ordered vacancies in NbN play a similar
role in the
electronic bandstructure to that observed in isostructural NbO and that
 the slight charge redistribution among
Nb and N atoms could be explained in terms of the structural anisotropy.
However, we  have calculational evidence that the observed new phase
cannot be a stoichiometric Pm3m phase of NbN.  It is energetically unfavorable
(by 1.00 eV/NbN) compared to Nb$_4$N$_4$ and the predicted lattice constant
is 5\% smaller than reported.  We are currently searching for an alternative
structure for the observed new phase.

The authors acknowledge Serdar \"{O}\u{g}\"{u}t and Karin M. Rabe
 for useful discussions.
E.C.E gratefully acknowledges the National Research
Council for financial support.

%
%

%

%
\begin{figure}
\caption{Crystal structure of Nb$_3$N$_3$. Large circles represent
 Nb sites, small circles represent N sites, and black dots
represent vacancy sites.  Atom types are interchangeable.
\label{crystal structure}}
\end{figure}
\begin{figure}
\caption{Self-consistent electronic band structure for (a) Nb$_4$N$_4$
(b) Nb$_3$N$_3$. The character of each band, using the method  described
 in the text, is indicated. Filled squares denote N p character,
filled circles denote Nb p character, and open circles denote Nb d character.
  The Fermi level is the energy zero.
\label{bandstructure}}
\end{figure}
\begin{figure}
\caption{Density of States (DOS) for (a) Nb$_3$N$_3$ and (b) Nb$_4$N$_4$
near the Fermi level.
For each phase, top panel shows total DOS, middle and bottom panels show
atom- and symmetry- projected partial DOS for nitrogen and niobium,
respectively.  Contributions from (Nb s and p, N s) states
not shown here are negligible.  The Fermi level is the energy zero.
\label{dos}}
\end{figure}
\begin{figure}
\caption{Theoretical equation of state for Nb$_4$N$_4$(open triangles)
 and Nb$_3$N$_3$ (open circles).  Energy is per NbN unit, with the Nb$_4$N$_4$
minimum taken as the energy zero.  The symbols are calculated values, while
the lines represent fits to the Birch equation.
\label{equation of state}}
\end{figure}
%
%
\begin{table}

\caption{Total and decomposed atomic charge distribution of Nb and N in
the Nb$_4$N$_4$ and Nb$_3$N$_3$ phases. }

\begin{tabular}{lcccc}
 &\multicolumn{2}{c} {Nb$_4$N$_4$} & \multicolumn{2}{c}{Nb$_3$N$_3$} \\
  &     \ Nb &    \    N     &\      Nb   & \    N    \\
\tableline
Q & \dec 40.17 &  \dec 7.83 & \dec 40.29 & \dec 7.71 \\
s & \dec 8.18   &  \dec 3.72 & \dec  8.15  & \dec  3.77 \\
p & \dec 18.15 &  \dec 4.11 & \dec 18.36 & \dec    3.94 \\
d    & \dec 13.84  & 0 & \dec 13.78 &    0 \\
\tableline
x   & & \dec 1.37 & & \dec   1.38 \\
y   & & \dec 1.37 &  & \dec   1.38 \\
z   &  &\dec 1.37 &  & \dec   1.18 \\
\tableline
xy  & \dec 2.88  & & \dec 2.62 & \\
yz & \dec  2.88  &  & \dec 2.82 & \\
zx  & \dec 2.88  &  & \dec 2.82 & \\
x$^2$-y$^2$&\dec 2.60  & & \dec 2.68 & \\
3z$^2$&\dec 2.60  & & \dec 2.84 & \\
\end{tabular}\label{table:charges}
\end{table}
\end{document}